# High Conductance Ratio in Molecular Optical Switching of Functionalized Nanoparticle Self-Assembled Nanodevices

Yannick Viero, Guillaume Copie, David Guérin, Christophe Krzeminski, Dominique Vuillaume, Stéphane Lenfant, and Fabrizio Cleri

Institut d'Electronique, Microélectronique et Nanotechnologie (IEMN), CNRS UMR 8520, and University of Lille, Avenue Poincaré, 59652 Villeneuve d'Ascq, France

**ABSTRACT:** Self-assembled functionalized nano particles are at the focus of a number of potential applications, in particular for molecular scale electronics devices. Here we perform experiments of self-assembly of 10 nm Au nano particles (NPs), functionalized by a dense layer of azobenzene−bithiophene (AzBT) molecules, with the aim of building a light-switchable device with memristive properties. We fabricate planar nanodevices consisting of NP self-assembled network (NPSANs) contacted by nanoelectrodes separated by interelectrode gaps ranging from 30 to 100 nm. We demonstrate the light-induced reversible switching of the electrical conductance in these AzBT−NPSANs with a record "on/off" conductance ratio up to 620, an average value of ca. 30 and with 85% of the devices having a ratio above 10. Molecular dynamics simulation of the structure and dynamics of the interface between molecular monolayers chemisorbed on the nano particle surface are performed and compared to the experimental findings. The properties of the contact interface are shown to be strongly correlated to the molecular conformation which in the case of AzBT molecules, can reversibly switched between a cis and a trans form by means of light irradiations of well-defined wavelength. Molecular dynamics simulations provide a microscopic explanation for the experimental observation of the reduction of the on/off current ratio between the two isomers, compared to experiments performed on flat self-assembled monolayers contacted by a conducting c-AFM tip.

## INTRODUCTION

With the ever increasing miniaturization of microelectronics, new paradigms for the fabrication and integration of electronic devices are becoming necessary.[1,2] Among the solutions proposed, a prominent one is the concept of molecular electronics, in which molecules can perform specific electronic functions reproducing all the features of microelectronic devices, such as amplification, commutation, storage and manipulation of data, and so on. Inspired by the original single-molecule concept of Aviram and Ratner,[3] a large variety of switchable molecules, that can be converted from one state to another by an external stimuli such as light, electricity or a chemical reaction has been synthesized (see a review in ref 4). Their performances are very well documented at the single-molecule level thanks to the emergence of powerful techniques such as break junctions (for a review, see ref 5). Despite their great promises, integrating single molecules into functional electronic devices at the nanoscale remains a big challenge.[6] In that trend, recent experiments started working on the implementation of two-dimensional (2D) assemblies of molecules arranged in a fully dense network, since this approach is more prone for device integration than single molecule experiments. This has been attempted by either depositing the molecules directly on a surface,[7−9] or by the intermediate of 2D-assembled nano particles (NP; typically Au, Ag, CdSe and the like) coated by a dense surface layer of molecules.[10−14] An even longer term objective of molecular electronics, still largely utopian for the moment, is to arrange the molecular network in such a way that it may imitate the operation of a living brain with its network of neurons.[15,16] This would be one possible strategy to evolve microelectronics from its current sequential operation mode, in which logic operations are processed one after another, to the parallel operation mode, in which the ensemble of logic gates work simultaneously on an ensemble of data.[10,15]

Here we report on the synthesis of self-assembled NP networks, in which each NP is functionalized by molecules whose conformation is switchable by means of irradiation under visible light. The purpose is that of obtaining a memristive behavior of the system, i.e., a device whose electrical resistance may change in a controllable and reversible way upon application of a voltage pulse, as already

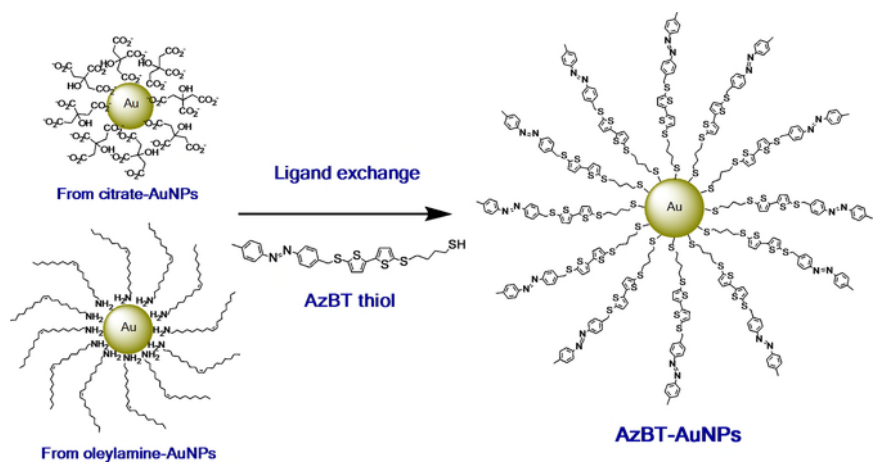

**Figure 1.** Synthesis of the 10 nm AzBT-capped AuNPs by two chemical routes.

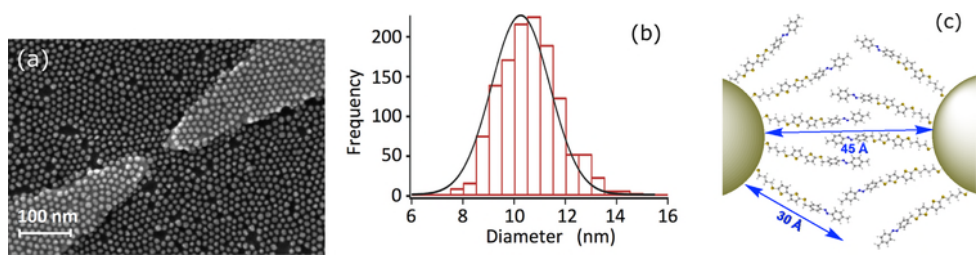

**Figure 2.** (a) Scanning electron microscope image showing the AzBT−AuNPs (citrate route) network deposited on a nanogap electrode (40 nm). (b) Statistical study of the Au nano particle size; average diameter is 10.2 nm, and average spacing is 4.5 nm. (c) Schematic of the interdigitation between molecules from adjacent Au nano particles.

demonstrated at a more macroscopic scale in hybrid NP/organic film synapstor (synapse-transistor).[17,18] Once coupled to the light-switchable behavior of the molecules, by adding a second input stimulus, the ultimate aim would be to build a device mimicking the learning capabilities of the brain, as shown for example in optically gated carbon nanotube transistors.[15,18−20]

Among the various types of molecules employed to coat the NP surfaces,[21−24] the azobenzene−bithiophene (AzBT) molecule contains the azobenzene group, which has the peculiarity of showing two different isomers both optically switchable (see Figure 1).[25,26] The two isomers differ by the value of the dihedral angle formed about the N≡N double bond, the more stable isomer (*trans*) having a dihedral of 180°, while the less stable (*cis*) is bent to a dihedral of about 10°.

It is possible to reversibly switch between the two AzBT isomers by illuminating the sample by light at the wavelength of 365 nm (*trans-to-cis* isomerization), and 480 nm (*cis-to-trans* isomerization), respectively. Recently, Smaali et al.[26] showed the possible utilization of this system as a molecular switch since the conductivity varies as a function of the isomerization state. In fact, the current (or "on/off") ratio between the *cis* and the *trans* isomer was determined to be ≈7000. (Note, however, that the identification of the *cis* state as the conducting one is not unambiguous: for example, Del Valle et al.[27] showed by first-principles calculations this to be the case for Si electrodes, while the opposite holding for carbon electrodes.) In that first study, the AzBT molecules formed a self-assembled monolayer covalently attached to a Au surface, and contacted by a conducting AFM tip representing the second electrode. However, this device configuration is not prone for large−scale integration. In the present work, we use instead the AzBT monolayer as the coating of 10 nm Au NPs, which are subsequently autoassembled forming a nanoparticles self-assembled network (NPSAN), on a surface carrying pairs of Au nano−electrodes separated by interelectrode gaps ranging from 30 to 100 nm (Figure 2). We demonstrate the light-induced reversible switching of the electrical conductance in these AzBT−NPSANs with a record "on/off" conductance ratio up to 620, an average value of about 30 and with 85% of the devices having a ratio above 10, which represents a strong improvement compared to previous reported results. Thus, as far as we know, the "on/off" ratio of photochromic NPSAN-based devices does not exceed 25 for one single switching event[14] or 1.5 for repetitive switching cycles.[13]

As it will be shown in the following, the conductivity measurements on such NPSANs show a striking phenomenon. The "on/off" ratio is much decreased in this configuration compared to the monolayer contacted by the conducting-AFM tip, going from the value 7000 down to about 600 in the best case. It is worth noting that such a decrease was also observed by other authors with a different kind of molecule.[13]

In the attempt of gaining a deeper insight into such phenomenon, we built an atomic-scale model of the NP−AzBT system and performed a series of large-scale empirical molecular dynamics (MD) simulations of the molecular interfaces, at ambient temperature and pressure. The MD simulation results provide a microscopic picture coherent with the experimental findings, as far as the structure of the NP−NP contacts is concerned. Moreover, the simulations also suggest possible explanations for the above experimental findings.

## MATERIALS AND METHODS

**Synthesis of 10 nm AzBT-Capped AuNPs.** The synthesis of the AzBT ligand bearing a terminal thiol function (AzBT–SH) was described in a previous work.[25] Two routes were used for the synthesis of AuNPs functionalized with this ligand (Figure 1). Both methods involved ligand exchange in the presence of AzBT–SH in 10 nm AuNPs preformed from two different precursors: citrate or oleylamine stabilized AuNPs. It was previously shown that these ligands can be substituted by thiols.[28,29] In the first route, the citrate–AuNPs (obtained in aqueous medium by the Turkevich method)[30,31] were precipitated in ethanol, centrifuged to eliminate the water supernatant, and then redispersed by sonication in the appropriate organic solvent for the subsequent thiolation step.[29] The citrate method has the advantage to supply AuNPs readily with very regular size. In the second route, we developed a synthesis of AzBT–AuNPs in organic medium starting from 10 nm oleylamine-capped AuNPs (see again Figure 1) following the protocol described by Wang et al.[32] Compared to citrate method, this route has the advantage to yield greater amount of material and more concentrated NPs solutions more suitable for the preparation of Langmuir film. More details on synthesis are given in the Supporting Information (Figure S1).

The thiolation of AuNPs with AzBT–SH was performed in organic medium for 3 days under inert atmosphere in the dark (see details in the Supporting Information). The ligand substitution and the switching capability of the AzBT–AuNPs was evidenced by UV–vis spectroscopy (see Supporting Information, Figure S2).

**Nanogap Electrodes Fabrication.** Coplanar nanogap electrodes were fabricated using standard electronic lithography processes. First, a ⟨100⟩ oriented silicon wafer was covered with a thermally grown, 220 nm thick, silicon dioxide, formed at 1100 °C during 135 min in a dry oxygen flow (2 L/min) and followed by a postoxidation annealing at 900 °C during 30 min under a nitrogen flow (2 L/min) in order to reduce the presence of defects into the oxide. Second, the e-beam lithography has been optimized by using a 45 nm-thick PMMA (3% 495 K), with an acceleration voltage of 100 keV and an optimized electron beam dose of 690 $\mu C/cm^2$ for the writing. After the conventional resist development (MIBK: IPA 1:3 during 1 min and rinsed with IPA), a metallic layer (1 nm of titanium and 10 nm of gold) were deposited by e-beam evaporation. Finally after the lift-off process using remover SVCTM14 during 5 h at 80 °C, well-defined coplanar electrodes separated by a gap comprised between 10 to 55 nm were realized (Figure S5 Supporting Information).

**Preparation and Deposition of the AzBT–AuNPs Monolayers.** A Langmuir film of AzBT–AuNPs was prepared following the method of Santhanam,[33] by evaporating a solution of NPs in TCE on the convex meniscus of a water surface in a Teflon Petri dish (see details in the Supporting Information, Figure S3). The transfer of the floating film was realized by dip coating directly on a lithographed substrate. Well-organized NPSANs were obtained by this method (see SEM image in Figure 2a). According to the statistical analysis of the NP size measured by SEM, the average diameter of the AzBT–AuNPs was 10.2 nm with a medium spacing of 4.5 nm within the network (Figure 2). It should be noted that the NPs size were more regular by the citrate method than by the oleylamine method (see Supporting Information, Figure S2).

Given the theoretical length of the free ligand (30 Å) with respect to the average spacing measured between the NPs (45 Å), one can assume that the AzBT molecules are partly interdigitated (see Figure 2 in the main text).

The chemical composition of the surface adsorbates were checked by X-ray photoelectron spectroscopy, a valuable tool to study the chemical composition of surface adsorbates (Supporting Information, Figure S4).

Ratios of experimental atomic concentrations S/C (0.15) and N/C (0.07) were in accordance with expected values (5/26 = 0.19 and 2/26 = 0.07, respectively). The N 1s and S 2p peaks appeared at 399.5 and 163.3 eV, respectively. These values are also very close to those obtained on AzBT SAMs. The S 2p signal was deconvoluted into a pair of doublets with a 1.2 eV splitting energy centered at 163.91 and 162.33 eV ascribed to sulfur bound to carbon (thiophene) and sulfur bound to gold, respectively. The characteristic binding energy of the $s_2p_{3/2}$ at 161.8 eV is in agreement with what is generally found for organothiols chemisorbed on Au.[33–35] This demonstrates that the grafting of AzBT–SH on AuNPs was successful. No sulfonate species was found at higher energy (around 168 eV) proving that no sulfur oxidation occurred. Following the method of Volkert,[36] we evaluated a ligand density of 2560 AzBT molecules per NP (3.2 molecules/nm$^2$) from the experimental S/Au ratio (0.357) of atomic concentrations.

**Electrical Measurements.** The electrical characteristics of nanogap electrodes covered with the AzBT–NPSANs were measured with an Agilent 4156C semiconductor parameter analyzer. The electrodes were contacted with a micromanipulator probe station (Suss Microtec PM-5) placed inside a glovebox (MBRAUN) with a strictly controlled nitrogen ambient (less than 1 ppm of water vapor and oxygen). Such a dry and clean atmosphere is required to avoid any degradation of the organics molecules. For the light exposure, an optical fiber was brought close to the nanogap electrodes inside the glovebox. For the blue light irradiation, we focused the light from a xenon lamp (150 W) to the optical fiber, and we used a dichroic filter centered at 480 nm with a bandwidth of 10 nm (ref 480FS10-50 from LOT Oriel). For the UV light irradiation, the optical fiber was coupled with a high power LED (ref M365F1 from Thorlabs). This LED has a wavelength centered at 365 nm and a bandwidth of 7.5 nm. At the output of the optical filter, the NPSANs were irradiated on about 1 cm$^2$ at power density of ∼0.1 mW/cm$^2$ (at 480 nm) and ∼6 mW/cm$^2$ (at 365 nm). The light illumination are done with no bias applied on the NPSANs, and then the current–voltage ($I$–$V$) curves are measured in the dark, thus avoiding any photo-current effect and decorrelating the electrical measurements from the illuminations.

**Computer Simulations.** We first performed energy-minimization calculations of the isolated AzBT molecular structure in vacuum with Gaussian-03,[37] using the density-functional theory in the local density approximation (LDA) with the B3LYP/6-311g functional. The geometries of the two ground states respectively for each isomer were optimized by a modified GDIIS algorithm implemented in Gaussian, and standard convergence parameters for the forces and displacements. The initial configurations were guessed using previously calculated, highly converged conformations of the *cis* and *trans* diazobenzene moiety. From such calculations, we also obtained the values of the point charges for all the atoms in the free molecule. We tested both the Mulliken and the Merz–Singh–

Kollman definition of the charges, finding but a small difference between the two definitions.

As in our previous studies of self-assembled molecular monolayers by classical molecular dynamics (MD),[38,39] we used the DLPOLY4 code[40] with the MM3 force field,[41,42] which includes bonding (stretching, bending, torsion, inversion) and nonbonding (van der Waals, Coulomb, and eventual H-bonding) force terms. The Au atomic lattice was first built by minimizing the fcc structure configuration with the Cleri−Rosato tight-binding potential,[43] and was kept frozen in the foregoing MD simulations, considering the very small lattice vibrations compared to the molecular motion at $T \leq 300$ K. Atomic trajectories in the constant-($NVT$) ensemble were calculated by the velocity−Verlet integrator, with a time step of 1 fs. Long-range electrostatic forces were evaluated by the Neumann reaction field method, as implemented in DLPOLY4.

Short-range van der Waals (VdW) interactions in the MM3 force field are described by a Buckingham-type two-body potential with a cutoff of 9 Å:

$$U(r_{ij}) = A \exp(-r_{ij}/\rho) - Cr_{ij}^{-6} \quad (1)$$

with $A$, $\rho$, and $C$ parameters depending on the atom species of $i$ and $j$ and $r_{ij}$ being the distance between the two atoms $i$ and $j$. The standard MM3 VdW parametrization was used also for the frozen Au surface.

Each molecule in the monolayer was anchored to one Au atom of the frozen via a thiol (S-bridge) bond, whose configuration and bond-stretching and bending parameters were taken from our previous work.[44] Both in the free-standing surface, and in the nanoparticle configurations, only the first one or two layers of surface Au atoms were explicitly represented. The surface number density of AZBT molecules is 3.6 mol nm$^{-2}$, as close as possible to the experimentally estimated value, compatibly with the requirements of maintaining a symmetric arrangement.

In the MM3 force field there are no parameters to describe the *cis* isomer as a stable state. Therefore, we had to construct an explicit dihedral angle C−N−N−C potential for the azobenzene group (the two C atoms being the C1 of each phenyl ring) of the type:

$$U(\phi) = A(1 + \cos \phi) + B(1 - \cos 2\phi) + C(1 + \cos 3\phi) \quad (2)$$

displaying a double minimum, the deeper one at $\phi = 180°$ describing the *trans* conformation and the shallower one at $\phi = 0$ describing the *cis*. However, we know from our DFT structure relaxation calculations that the dihedral for the *cis* cannot be exactly zero, because of the hindrance coming from the hydrogen atoms, but rather $\phi \approx 10°$. In practice, it was not possible to obtain a stable empirical potential of the form above, for such values of $\phi$. The best compromise was reached with $A = 0$, $B = 3.75$, $C = 7.5$, for which the dihedral fluctuates around $\phi \approx 38°$ and is stable up to simulation temperatures of about 300 K. The AzBT molecules have, however, the tendency to go back to the *trans* conformation even with this potential, at the higher temperatures during a MD simulation. For this reason we also added harmonic restraints during the MD simulations with all-*cis* molecules.

## RESULTS

**Electrical Transport in NPSANs.** Most of the electrical results reported in this section have been obtained with NPs synthesized by the oleylamine route, unless specified. As mentioned in section 2, the main reason is that this route, compared to citrate method, has the advantage to yield greater amount of material and more concentrated NPs solutions more suitable for the preparation of Langmuir film. Moreover, we have observed that the electrical transport properties are the same for AzBT−NPSANs fabricated by the two methods.

Figure 3 shows the typical I−V curves measured on a NPSAN with a nanogap distance of 73 nm (SEM image in

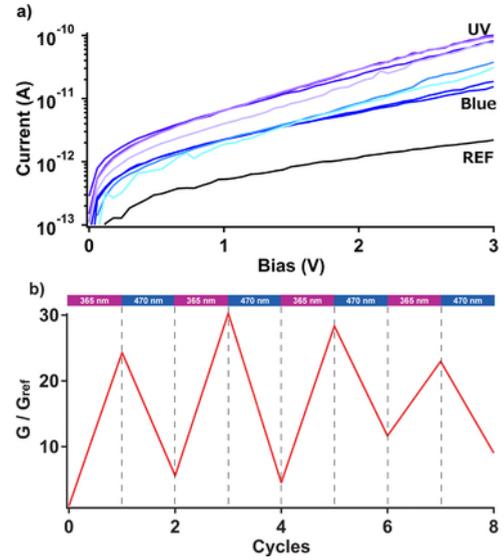

**Figure 3.** Reversible photoswitching of the conductance of an AzBT−NPSAN nanodevice. (a) Typical current−voltage (I−V) curves measured before any illumination (dark curve) and after each illumination at 365 nm (purple curves) and 470 nm (blue curves). (b) Plot of the normalized conductance $G/G_{ref}$ (current taken at 3 V; $G_{ref}$ is about $7 \times 10^{-13}$ A, from the "REF" curve in part a), showing that the UV illumination brings the device into a conductive "on" state, whereas a blue illumination brings it to an "off" state.

Figure 2a). These data demonstrate the reversible photoswitching of the conductance of the NPSAN at room temperature. Before any illumination, we measure the lowest conductance of the nanodevice (black curve in Figure 3a). After UV illumination at 365 nm for 1 h (trans-to-cis isomerization), the current (measured in the dark) through the NPSAN is significantly increased by a factor of about 25 (at +3 V). Note that this "on state" current saturate after 1h of illumination (no further increase for longer time exposure), which means that almost all the AzBT molecules have switched in their cis configuration.

Subsequently, when illuminating the NPSAN by blue light (480 nm for 1 h, cis-to-trans isomerization of AzBT) and then measuring again the I−V curve in the dark, we observe a decrease of the current. However, the current does not return to the initial value. If we consider the same photoiosomeriza-tion cross section for the trans-to-cis and cis-to-trans isomer-ization as shown in our previous work on self-assembled monolayer (SAM) molecular junction with the same AzBT molecules (AzBT−SAM),[25,26] this may be simply explained by the lower photon flux under blue light than under UV light (see section 2.4 above). If we assume a simple exponential decay of the current under blue light illumination (as shown in ref 26), we can extract, from the current ratio between the blue and UV illumination, a characteristic time constant for the cis-to-trans isomerization, $\tau_{CT}$, of about 37 min. This value is in good agreement with our previous determination of 20 min measured from switching kinetics experiments on SAMs of AzBT (at a light power density of 0.25 mW/cm$^2$ in this latter case, instead of 0.1 mW/cm$^2$ in the present work). Note that we cannot exclude some molecular rearrangement of the AzBT molecules around the NPs in the network under the first trans-to-cis isomerization, which may impede further complete back isomerization to the trans state.

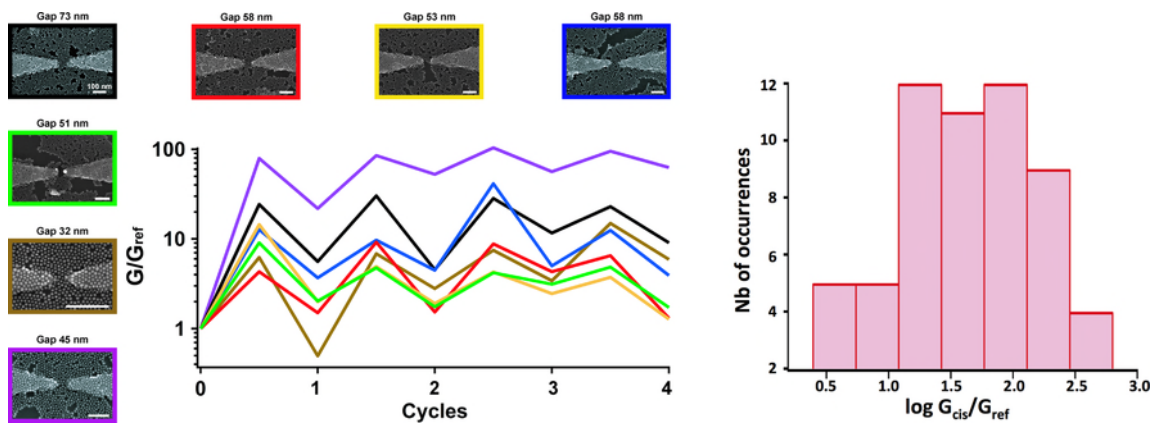

**Figure 4.** (a) Conductance switching of 7 AzBT NPSANs for four complete UV/vis illumination cycles. Insets are SEM pictures of the devices taken at the end of the experiments. The colors surrounding the insets are associated with the corresponding colors of the curves. (b) Histogram of the "on/off" current ratio for 62 devices (10 synthesis via the citrate route, and 52 via the oleylamine route). The ratio spans the range 3−620, exceeding 10 for more than 85% of the devices.

We repeated this illumination cycle several times and the different I−V curves are shown in Figure 3a, demonstrating a good reproducibility. Note also that the cis state in these NPSANs is stable for a long period of time (days)—no spontaneous cis-to-trans thermal isomerization at room temperature—also in agreement with our previous measurements on SAMs.[26]

Figure 3b shows the corresponding value of the normalized conductance ratio $G/G_{ref}$ (measured at 3 V), with $G_{ref}$ the conductance measured on the pristine NPSAN corresponding to the "off state". We clearly observe on/off conductance ratios between 25 and 30. For the sake of completeness, we performed the same experiment with alkanethiols-NPSAN-based nanodevices and did not measure any change in the conductance of the device (not shown).

To test the device-to-device reproducibility, we repeated the same experiment for numerous devices. Figure 4a shows "on/off" conductance ratios similar to 3b, for 7 devices with nanogaps ranging from 32 to 73 nm (SEM images shown as insets). The $G/G_{ref}$ ratios are always calculated at the maximum applied bias, $V_{max}$, of the I−V scans, which spans from 1 to 6 V depending on the device. This maximum applied bias is set to obtain a measured current in the 0.05−5 nA range at $V_{max}$, such a large dispersion depending on the nanogap size, the number of NPs in the NPSAN, and the molecular arrangement in the network. While we did not observe a clear scaling of the current level with the geometrical NPSAN parameters, but a rather large device-to-device dispersion, Figure 4a shows that the photoswitching behavior is reproducible for at least 4 complete illumination cycles, with "on/off" conductance ratios up to 100.

Focusing on the first switching event (trans-to-cis isomerization under UV light), in Figure 4b we show the histogram of the "on/off" ratio for 62 devices. This ratio is systematically greater than 3 with a huge majority (85%) greater than 10, 25% greater than 100, up to a maximum of 620. A fit by a log-normal distribution give a log-mean of 1.53, i.e., a mean ratio of 34.

The switching behavior of the AzBT−NPSANs is therefore consistent with that of AzBT−SAM junctions,[25,26] i.e. a larger current ("on" state) with AzBT in the cis configuration than in the trans one ("off" state). However, albeit the on/off ratio of the AzBT−NPSANs (up to 620) is much larger than those reported for several other NPSAN (using diaryethene, on/off ratios of 1.8,[13] and 25,[14] have been reported), it remains lower than the one measured on AzBT−SAM devices, up to $7 \times 10^3$.[25] This peculiar feature will be discussed later, after considering the theoretical results given below in the next sections.

**DFT Calculations on the Isolated Molecules.** The optimized geometrical conformations of the two isomers (cis and trans) of the AzBT molecule mainly differ by the value of the dihedral angle of the double N≡N bond, which decreases from 180° down to 10.64°. The total length of the molecule defined by the distance between the two utmost hydrogen atoms is strongly influenced by such a bending, as it decreases from 28.6 (trans) down to 23.5 Å (cis).

The electronic properties of the neutral and charged states of the molecule (±1e) have also be calculated. Table 1 summarizes the main results obtained for the different cis/trans AzBT configurations. Relatively modest variations are observed in terms of electronic properties. The same shift of the ionization potential for two isomers of the isolated diazobenzene molecule (used as a reference) is observed, however with a modest amplitude, due to a higher delocalization of the π state on the aromatic part of the AzBT molecule compared to the diazobenzene.

**Table 1**

| molecule | affinity (eV) | ionization (eV) | gap (eV) | dipole (Debye) | length (Å) |
|---|---|---|---|---|---|
| AzBT (cis) | −1.28 | −6.95 | 3.77 | 8.01 | 23.5 |
| AzBT (trans) | −1.46 | −7.07 | 3.37 | 5.15 | 28.6 |
| diazobenzene (cis) | −0.60 | −7.66 | 3.74 | 3.75 | 9.15 |
| diazobenzene (trans) | −0.88 | −8.17 | 3.53 | 0 | 11.27 |

Next, the optical properties have been investigated by TDDFT (time-dependent density functional theory) as implemented in Gaussian03. A plot of the theoretical absorbance for the two AzBT isomers is given in the Supporting Information, Figure S6. A major active transition at 353 nm is predicted for the *trans* isomer, which can be linked mainly to the HOMO−1 → LUMO (50%) transition. For the *cis* isomer, however, the transition at 337 nm could be linked to the HOMO → LUMO+1 transition (84%) with a oscillator strength reduced by 50% with respect to the optical transition near 300 nm. A comparison with the experimental spectra (see Supporting Information, Figure S2) confirms the predicted reduction of absorbance close to 340−350 nm, during the transition from the *trans* to the *cis* isomer.

**MD Simulation of Spherical Nanoparticles.** In a first series of MD simulations, we considered two perfectly spherical NP surfaces, facing each other at a variable distance $d$ along the common diametral line. The 25 AzBT molecules attached on each of the NP surfaces were bonded to Au atoms regularly spaced on the lattice sites of a curved fcc (111) surface with spherical radius of 10 nm. We compared the NP configurations with all the molecules on each side either in the *trans*, or in the *cis* conformation. MD simulations were performed at constant-(NVT), for temperatures $T = 50, 100, 200, 300$ K controlled by a Langevin stochastic thermostat. Each simulation for a given NP−NP distance $d$ was initially equilibrated for 100 ps, and subsequently statistical averages were calculated over 1 ns ($10^6$ time steps). The minimum of the total energy is obtained for $d = 4.5$ nm, in good agreement with the experimental observations above.

At the lowest temperatures $T \leq 100$ K, the molecules in *trans* configuration are nicely interdigitated (see snapshots in Figure 5), while the molecules in the *cis* configuration have (at least visually) a smaller contact density. By increasing the temperature, at $T > 200$ K the molecules in the *cis* configuration completely lose contact with each other, and at even higher temperatures the molecular monolayers tend to collapse on the respective NP surfaces. Conversely, still at $T = 300$ K the *trans* configuration still displays some contact.

In order to better characterize the intermolecular interaction, essentially due to VdW forces between the aromatic rings of facing molecules, we have calculated the dynamic average of the angle Ψ between the planes containing the azobenzene moiety of any two adjacent molecules (see scheme on top of Figure 5). We note that Ψ = 0 means parallel, and Ψ = π/2 perpendicular, for the *trans* isomer; however, for the nonplanar structure of the *cis* (see above), a relative angle of Ψ ≈ π/2 results in the terminal phenyls from facing molecules to be rather parallel, while Ψ ≈ 0 results in almost perpendicular phenyl rings, i.e. quite the opposite of the *trans* configuration. Notably, the more two aromatic rings from adjacent molecules are parallel to each other, the more favorable is the π−π stacking interaction. In practice, we calculated the value of the angle Ψ for any pair of molecules, whose N≡N bond centers were separated by ≤1.2 nm, with the condition that each molecule belongs to a different NP; statistics were accumulated every 1 ps, during the 1 ns MD simulation.

The time-dependent plot in Figure 5 shows the evolution of the molecule-averaged ⟨Ψ⟩ at $T = 100$ and 300 K, for the two isomers. At low temperature we find ⟨Ψ⟩ ≈ 50° for both isomers, however with larger fluctuations for the *cis* because of the lower interaction between the molecules. Because of the above definition, such a value means that the *cis* tends to be somewhat closer to parallel orientation (i.e., more π−π interaction), than the *trans*. However, bending into the *cis* form also shortens the molecule end-to-end length by about 0.3 nm compared to the fully elongated *trans* form, as well as creating a bigger steric hindrance in the lateral direction with respect to the molecule axis. Such factors conjure toward a reduction of the intermolecular interaction, despite the slightly better degree of parallelism.

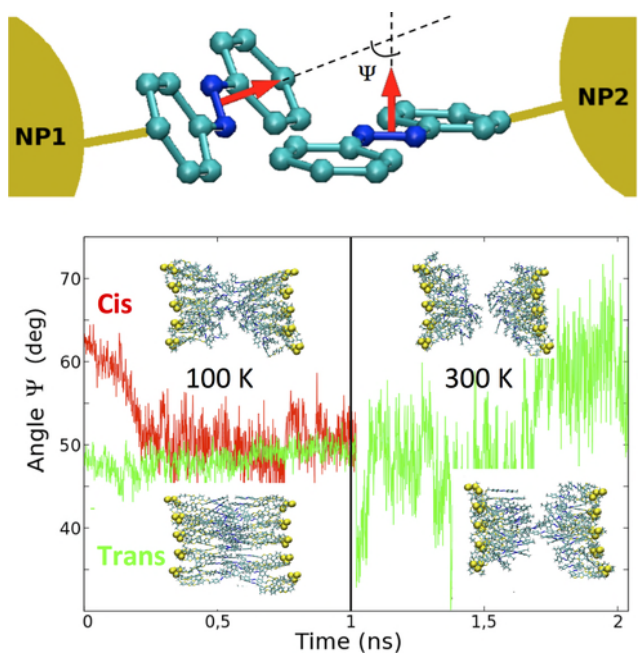

**Figure 5.** Above: Schematic of the molecule-averaged angle ⟨Ψ⟩, formed by two azobenzene moieties from two AzBT molecules bonded to the two facing nano particles. The red arrows represent the direction of the normal to the plane containing the azobenzene. Below: time evolution of the average value ⟨Ψ(t)⟩ during two 1 ns MD simulations at $T = 100$ and 300 K. Key: red curve, *cis*; green, *trans* isomer. Also shown are the snapshots of the final configurations of the AzBT−coated NPs, in the *cis* (above) and *trans* (below) conformations. At $T = 300$ K, the *cis* molecules are completely disconnected, and the corresponding fluctuation plot of the angle is no longer meaningful.

At higher temperature (Figure 5, right side of the ⟨Ψ(t)⟩ plot), the fluctuations of ⟨Ψ⟩ about its mean value become much larger for the *trans*, while the *cis* have completely lost contact with each other, and any meaningful π−π interaction can no longer be detected.

**MD Simulation of Faceted Nanoparticles.** For a size of about 10 nm, the shape of Au NPs should however not be a perfect sphere,[45] but a rather regular surface, predominantly {111}-faceted, conducting to icosahedral- or octahedral-like shapes.

To evaluate the influence of the NP shape on the intermolecular interaction, we then built two (111) Au facets, each composed of two hexagonal portions of surface with side 2 nm and area 9 nm$^2$, placed at a relative distance $d = 4.5$ nm. On the two facets, 20 and 17 AzBT molecules were respectively bonded, with a constant average surface density, and positioned so as to be easily interdigitated at $T = 0$K. As shown in Figure 6a,b, the interaction between molecules from nearby NPs survives also at the highest temperature $T = 300$ K, for both isomers. From a purely geometrical point of view, the number of interacting molecules on the portion of planar surface is larger than for a spherical surface, which explains the increased stability of this system.

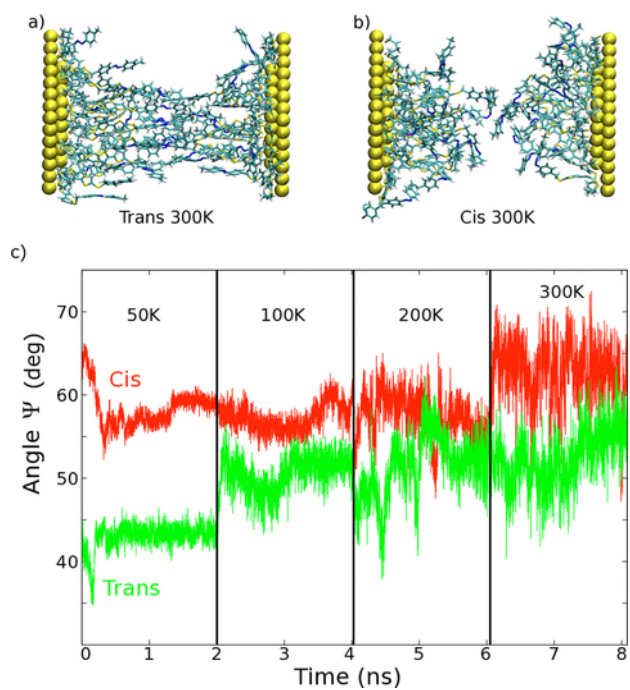

Figure 6. Above: Snapshots of the final configurations of the AzBT *trans* (a), and *cis* isomers (b), bonded on the two parallel Au 111 facets at a relative distance $d$ = 4.5 nm, after a 2 ns MD simulation at constant-(NVT) with $T$ = 300 K. Below (c): Plot of the average value of the angle $\Psi$ during four 2 ns MD simulations at $T$ = 50, 100, 200, and 300 K. Key: red curve, *cis*; green, *trans* isomer.

The plot of the average angle $\langle\Psi\rangle$ between the azobenzene planes of adjoining molecules is shown in Figure 6c for the two isomers, at different temperatures of the MD simulation. The molecule-averaged $\langle\Psi\rangle$ in this case is consistently larger for the *cis* than for the *trans* isomer at all temperatures. At $T$ = 100 K, the *cis* has $\langle\Psi\rangle$ = 58°, compared to $\langle\Psi\rangle$ = 53° for the *trans* form, and both are larger than the $\langle\Psi\rangle$ = 50° for the spherical NP surface. The two values appear to merge to a common value toward the highest temperatures, with obviously larger thermal fluctuations.

By looking at the periphery of the hexagonal facets, we can observe an evident tendency to folding of the AzBT molecules toward the surface of the respective NP facet, an effect that increases upon increasing the temperature. This can be explained by the combination of two cooperative factors: the VdW attraction from the Au surface, and a reduced interaction between parallel molecules around the periphery, compared to the molecules in the middle of the facets. While the VdW attraction per unit surface should not change, the nearest-neighbor interaction between parallel molecules clearly depends on the particular geometry we chose for the system. It is therefore possible that the observed folding is an artifact of the simulation, and that in the presence of a more compact 2D layer of self-assembled NPs such phenomenon disappears, leading to a better contact between NPs.

**MD Simulation of Dense Nanoparticles.** In order to represent a more realistic configuration, we finally designed compact, 2D-periodic arrangements of Au NP with faceted shapes. Figure 7 represents the NP assembly, for the case of perfectly icosahedral NPs with average diameter 10 nm. In order to reduce the complexity of the system, we included the AzBT molecular monolayer only in the contact region between the three NPs in the 2D-periodic structure. 143 AzBT molecules were placed on each NP, for a total of 429 molecules (corresponding to about 25 000 atoms). As before, only the Au atoms defining the surface of the 111 facets are explicitly included and frozen, their role being only of supplying fixed bonding sites for the terminal S atom of each AzBT molecule, and to provide the surface VdW interaction.

The interface between any two icosahedral facets is no longer parallel (Figure 7b), as it was the case for the simulations of the previous sections. Therefore, the "optimal" NP−NP average distance $d$ = 4.5 nm could indeed correspond to quite different separations of the facets. We therefore performed different sets of MD simulations, by freezing the Au NPs at three different distances, where the 4.5 nm was imposed alternatively at the three positions indicated in Figure 7b, thus leading to different effective NP−NP separations in each case. As before, we performed several sets of MD simulations in the constant-(NVT) ensemble, at temperatures ranging from 50 to 300 K.

Case 1 (i.e., distance of 4.5 nm between the upper, closest edges of icosahedral NP facets) corresponds in fact to the *largest* average NP−NP separation $d$. The MD simulations resulted in a net separation of the interfaces for the *cis* conformation, already at very low temperatures, therefore the corresponding simulations will not be further shown in the following for either isomer. The top panel in Figure 8 shows the snapshots of the final configurations for the case 2 (4.5 nm distance between the centers of the NP facets) and case 3 (4.5 nm distance between the lower, farmost edges of the NP facets, corresponding to the *closest* average NP−NP distance $d$), for both the *trans* and *cis* isomeric state of the AzBT molecules. For clarity, molecules on the NP surfaces are represented only by dots, indicating the center of the terminal group.

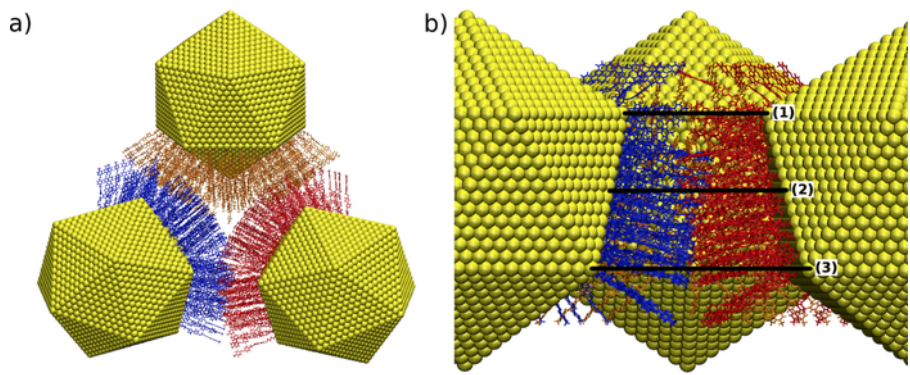

Figure 7. Schematic of the system comprising three Au nano particles with icosahedral shape, functionalized by a monolayer of AzBT molecules. The yellow spheres represent Au surface atoms. (a) 2D view from above. (b) View in the plane. The lines labeled (1), (2), and (3) in part b show the three possibilities for measuring the relative NP−NP distance.

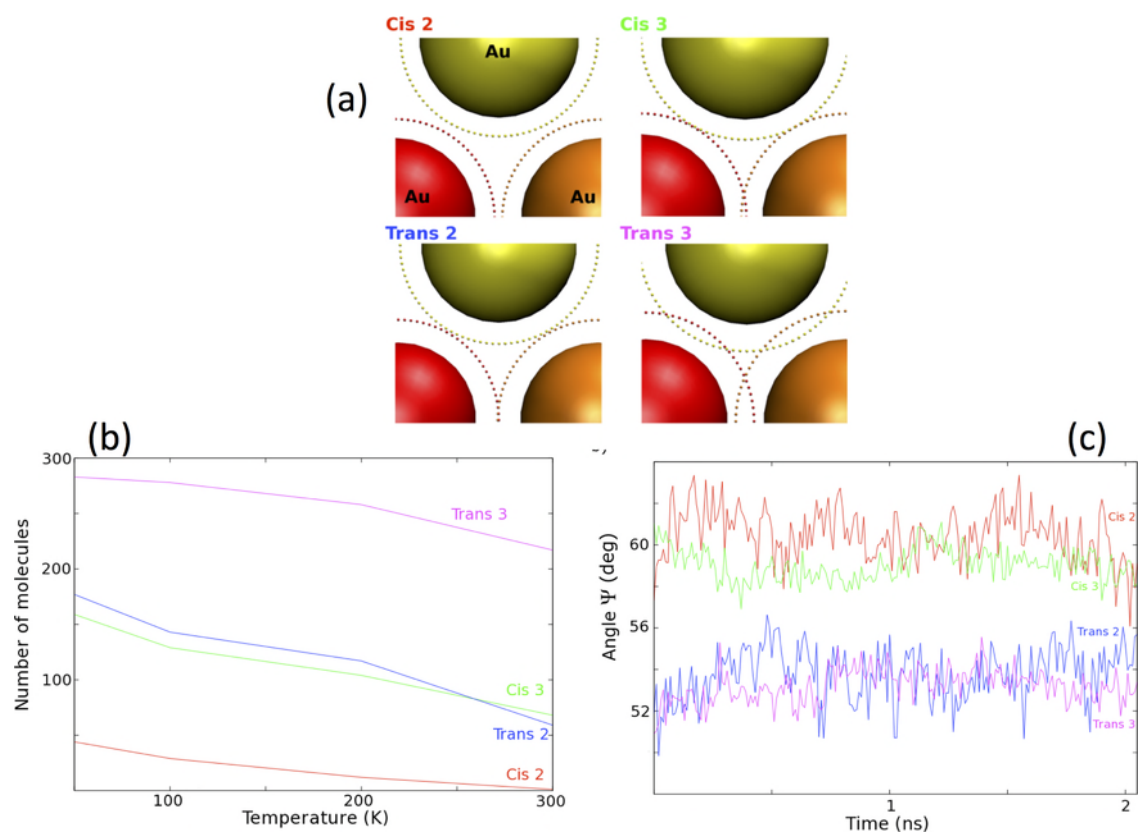

**Figure 8.** (a) Average configuration of the molecular monolayers coating the three adjoining Au nanoparticles, after 2 ns of constant-($NVT$) MD simulation, for the four systems studied (cases 2, smaller NP−NP separation, and 3, larger NP−NP separation). Small dots represent the center of each molecule terminal group. Both the fraction of contact surface, and the interdigitation of the molecules, appear to be increased for the *trans* isomer compared to the *cis*. (b) Plot of the average number of interdigitated molecules as a function of the temperature. (c) Plot of the average value of the angle Ψ for 2 ns constant-($NVT$) MD simulations at $T$ = 300 K. Color curves in parts b and c correspond to color labels in panel a.

For each of these four sets of MD simulations, an estimate of the degree of interdigitation of the molecules from opposing NP surfaces was carried out. If $L$ is the end-to-end length of one molecule, interdigitation is possible whenever $d < 2L$. For practical purposes, two molecules from adjacent NP were considered interdigitated if their terminal atoms are closer to the opposite NP surface than to their own NP. Figure 8b displays a plot of the average number of interdigitated molecules as a function of the increasing temperature. For all cases, it is seen that the degree of interdigitation decreases linearly with $T$, corresponding to the progressive activation of the torsion and bending degrees of freedom of the molecules, thus allowing wider and independent deformations, and therefore less superposition. Furthermore, for a given NP− NP effective separation, the degree of interdigitation is always better for the *trans* isomer, which can be explained both with the slightly longer end-to-end length of the molecule in this conformation, and by the reduced steric lateral hindrance, both effects allowing a better superposition among AzBT molecules coming from adjacent NP surfaces.

We calculated as well the molecule-averaged $\langle\Psi\rangle$, plotted in Figure 8c for the four different systems. The same qualitative difference between the *cis* and *trans* isomers as before could be observed, with a slightly larger value of $\langle\Psi\rangle$ = 5 9 ° of the former, compared to $\langle\Psi\rangle$ = 5 3 ° for the latter, at the highest temperature $T$ = 300 K. However, we note that in such a complex configuration, some molecules always go back to the *trans* conformation in the all-*cis* simulations, therefore the single average value of $\langle\Psi\rangle$ is likely a less representative quantity here. Also, it is worth noting that the relative NP−NP separation (blue-red vs green-purple sets of curves in the figure) does not seem to significantly influence $\langle\Psi\rangle$.

## DISCUSSION

The extensive DFT and molecular dynamics simulations described in the previous section attempted at modeling the dynamical structure of the interfaces between Au nanoparticles functionalized by a dense monolayer of azobenzene-bithiophene molecules.

The AzBT molecule has two stable isomers: the *trans*, fully elongated with the azobenzene moiety lying flat in the plane of the molecular axis, and the *cis*, with the azobenzene moiety bent to a dihedral angle of about 10° and shorter by about 0.3 nm than the former, as shown by structure optimization. DFT calculations on the isolated isomers in vacuum confirmed the variation of optical absorption properties during the *trans/cis* switching. However, they also indicate that relatively modest changes are expected in terms of electronic properties. Therefore, only electronic structure effects cannot reasonably explain the huge variation of the "on/off" ratio observed for the SAM based system. On the other hand, DFT structure relaxation calculations indicate a substantial change of molecule length, which, in the framework of a nonresonant tunneling transport through the molecule, would have a major influence on the overall electrical barrier, and suggest a much larger impact of the geometrical effects, both in terms of electronic coupling and effective tunneling distance.

In general terms, our MD simulations of NP surfaces covered by a monolayer of AzBT molecules, showed that the *trans* isomer typically gives rise to a more stable NP−NP interface structure, thereby increasing the intermolecular interactions and, notably, the degree of π−π stacking between molecules from adjoining NP surfaces. Note that our simulations cannot access the conductivity per se, but give structural indications, from which the behavior of conductivity may be inferred, at least in part. Therefore, this is only a relative effect, which may act on the reduction of the *cis/trans* (or "on/off") ratio: we can only state that the numerator of the ratio is more severely affected than the denominator, by this ensemble of structural modifications (but from this, it would be impossible to deduce an absolute value of the ratio).

In the light of these results, the reduction of the "on/off" ratio experimentally observed in the NP self-assembled networks, compared to that for the molecular monolayer contacted by an AFM tip,[26] can be explained by two concurrent phenomena. The first one concerns the extent of the contact surface between adjacent NPs, which is consistently larger for the *trans* conformation of the AzBT molecules at the average NP−NP distance $d$ = 4.5 nm: under such conditions, electrons have a larger tunnelling probability to jump between NPs. The second one is the much higher degree of interdigitation for the *trans* conformation, contributing a better π−π interaction between adjacent molecules. In fact, for the conductivity measurements on the monolayer, the conducting AFM tip contacts the molecules by its edge: in this way, the π-orbitals of the *trans* isomer do not interact with the tip, while in the *cis* conformation the same orbitals are parallel to the tip. By contrast, in the case of the contact between NPs a better interdigitation implies a more substantial π−π interaction, and therefore a higher conductance (higher probability of tunneling).

The sum of these two effects (larger contact surface + interdigitation), make for an improved conductivity of the *trans* form (the "denominator" of the ratio) compared to the *cis* (the numerator), thereby reducing the "on/off" ratio as experimentally observed. For instance, Wu et al.[46] (while their experiment is not directly and quantitatively comparable to the present work) have shown that π−π interactions increase the conductance of mechanically controlled break junctions with aromatic molecules. Moreover, it is known from theoretical calculations that decreasing the distance between two π−π interacting conjugated molecules by 0.5 Å increases the transfer integral (i.e., electronic coupling) by a factor between 3 and 4,[47] which has a strong impact on the electron transport properties. Finally, we note that successive switching can induce some strain effect and some reorganization of the nanoparticles (as well as the interconnecting molecular network) in the NPSANs. This feature may be responsible for the "fatigue" effect (small, progressive decrease of the "on/off" ratio, upon successive trans/cis switching), sometimes observed in our measurements.

## CONCLUSIONS

In summary, we performed a combined experimental and theoretical study of the conductivity of self-assembled 2D networks of Au nanoparticles functionalized by a dense monolayer of azobenzene-bithiophene molecules. Such a system displays two different isomeric states, namely a highly conducting one with all the molecules in the *cis* conformation, and a less conducting one with the molecules in the *trans* conformation. The molecules can reversibly switch between the two states, by a sequence of light irradiations at different wavelengths. We fabricated planar nanodevices consisting of NP self-assembled network contacted by nanoelectrodes separated by interelectrode gaps ranging from 30 to 100 nm. We demonstrated the light-induced reversible switching of the electrical conductance in these AzBT−NPSANs with a record "on/off" conductance ratio up to 620, an average value of about 30, and with 85% of the devices having a ratio above 10. However, experimental measurements of the tunnelling current showed a much reduced "on/off" ratio, compared to previously measured values of up to $7 \times 10^3$.[26]

In order to provide a microscopic explanation to such observations, we built atomistic models of the interfaces between Au nano particles coated by dense layers of AzBT molecules, and carried out molecular dynamics simulation of such structures at finite temperature. The results of the MD simulations demonstrate a much better stability of the interfaces formed by the *trans* isomer, as well as an increased intermolecular interaction in such conformation. These results allowed to formulate a microscopic explanation for the observed experimental results, namely the decrease of the *cis/trans* current ratio is understood as originating from the improvement of the conductivity of the *trans* isomer, compared to the conducting-AFM experiments on monolayers.[26] Such an improvement is related to the better contact surface, as well as the better interdigitation of the *trans* molecules from close nanoparticles. Overall, this work shows that NPSAN devices can be envisioned as functional molecular-based switches at the nanoscale.

## ASSOCIATED CONTENT

### Supporting Information

The Supporting Information is available free of charge on the ACS Publications website at DOI: 10.1021/acs.jpcc.5b05839.

(1) Synthesis of citrate Au nanoparticles and their transfer in organic medium; (2) synthesis of the oleylamine-capped Au nanoparticles, with UV spectra and SEM images; (3) synthesis of AzBT-capped Au nanoparticles, with UV−vis spectra; (4) preparation of nanoparticle self-assembled networks (NPSAN), with XPS spectra and peak attribution; (5) fabrication of nanogap electrodes with SEM images; and (6) the ab initio (Gaussian-03) calculation of the theoretical absorption spectrum of the free AzBT molecule, in the two *trans* and *cis* states (PDF)

## AUTHOR INFORMATION

### Corresponding Authors

*(S.L.) E-mail: stephane.lenfant@iemn.univ-lille1.fr.
*(F.C.) E-mail: fabrizio.cleri@univ-lille1.fr.

**Notes**

The authors declare no competing financial interest.

## ACKNOWLEDGMENTS

Y.V. and G.C. contributed equally to this work. Work was financially supported by the EU FET Project 318597 (SYMONE: Synaptic Molecular Networks for Bioinspired Information Processing) and by the French ANR Project ANR-12-BS03-2012 (SYNAPTOR). Computing resources were provided by the French National Supercomputing Center CINES under Project c2014-097224.

# High Conductance Ratio in Molecular Optical Switching of Functionalized Nanoparticle Self-Assembled Nanodevices.

**Yannick Viero, Guillaume Copie, David Guérin, Christophe Krzeminski, Dominique Vuillaume, Stéphane Lenfant, and Fabrizio Cleri.**

*Institute for Electronics Microelectronics and Nanotechnology (IEMN), CNRS and University of Lille, Av. Poincaré, 59652 Villeneuve d'Ascq, France.*

## SUPPORTING INFORMATION

### *Synthesis of citrate-AuNPs and transfer in organic medium.*

We followed the method of Bernard et al.[1] All chemicals were purchased from Aldrich (except oleylamine 95% supplied by Strem Chemicals) and they were used without further purification. Citrate-AuNPs were synthesized by the Turkevich method.[2-3] To obtain a 100 mL aqueous solution of AuNPs, a solution with 1 mL of tetrachloroauric acid trihydrate $HAuCl_4.3H_2O$ (1%) in 79 mL of deionized water was first prepared. A 20 mL reducing solution with 4 mL of trisodium citrate dihydrate (1%) and 80 μL of tannic acid (1%) in 16 mL of deionized water was then added rapidly to the Au solution under vigorous stirring (all solutions at 60 °C). The mixture was boiled for 10 min before being cooled down to room temperature. A continuous stirring was applied throughout the process. Transfer of citrate-AuNPs in organic medium was necessary for the thiolation reaction. To this end, the 100 mL citrate-AuNPs solution was centrifuged at 10000 tr/min for 30 min to eliminate the maximum of water supernatant. NPs were precipitated by addition of ethanol, then again centrifuged at 7000 tr/min for 5 min. The black precipitate was redispersed in 5 mL of DMF by sonication, providing a dark blue solution immediately used for the thiolation step with AzBT-SH.

### Synthesis of the oleylamine capped AuNPs.[4]

In a schlenk flask equipped with a nitrogen inlet and a condenser was dissolved 50 mg of $HAuCl_4.3H_2O$ in 5 ml of oleylamine (95%) and 5 ml of anhydrous toluene. Under nitrogen atmosphere, the homogenous orange solution was maintained at 80°C under stirring overnight. During this time, the color changed from orange to dark red. AuNPs purification was performed by repeated centrifugation/sonication cycles of the solution: the reaction mixture was first diluted with 50 % of hexane, followed by addition of ethanol to precipitate the NPs. After centrifugation for 10 min at 7000 tr/min, the supernatant was eliminated, then the NPs were redispersed in hexane by sonication. This process was repeated three times and finally the oleylamine-AuNPs were redissolved in 5 mL of toluene (good stability in this solvent several months at 5°C). By the same precipitation/centrifugation process, toluene was easily replaced by other solvents for ligand exchange with AzBT-SH. The surface plasmon resonance peak (SPR) was observed at 524 nm in $CHCl_3$ by UV-vis spectroscopy. Average diameter of AuNPs measured by the statistical analysis of SEM image was 9.4 nm (Figure S1).

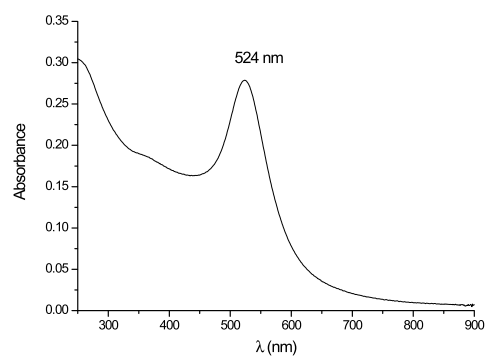

(a)

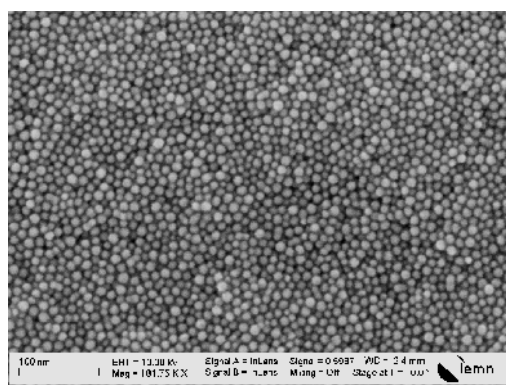

(b)

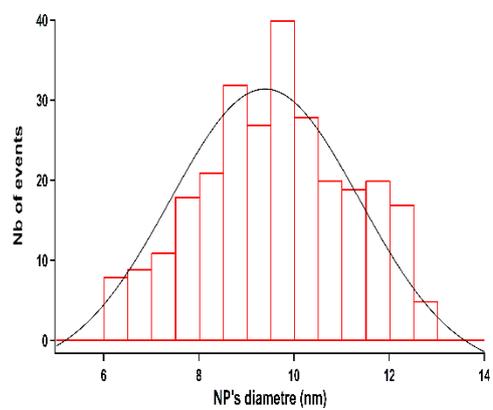

(c)

***Figure S1 :*** *(a) UV-vis spectrum of oleylamine-AuNPs in CHCl$_3$, (b) SEM image of a NP monolayer and (c) the corresponding average diameter by SEM in an evaporated drop.*

## Synthesis of AzBT-AuNPs.

All solvents were perfectly degased by nitrogen bubbling. Dimethylformamide (DMF), dimethylsulfoxyde (DMSO) or tetrachloroethane (TCE) turned out suitable solvents both for free ligands and functionalized-AuNPs. In a nitrogen glovebox ($O_2$ and $H_2O$ < 5 ppm), 5 mg of AzBT-SH (synthesis described in ref. [5]) were added to 1 mL of the previous oleylamine or citrate-AuNPs in organic medium. The thiolation was performed at room temperature for 3 days under nitrogen in the dark. The ligand excess was eliminated as following: the reaction mixture was first diluted with 50 % of dichloromethane (solvent in which AzBT-AuNPs are few soluble). After centrifugation for 3 min at 7000 tr/min, the supernatant was eliminated then the precipitate was cleaned thoroughly by fresh $CH_2Cl_2$ (without sonication to avoid aggregation). Finally, a concentrated red-purple solution of AzBT-AuNPs was obtained by sonication in 1 mL of TCE. We observed a good stability of this solution over long period at 5°C (several months). AzBT-AuNPs average diameter measured by SEM was 10.2 nm.

## UV-vis spectroscopy

Figure S2 compares UV-vis spectra of oleylamine-AuNPs and AzBT-AuNPs in DMF. Only the surface plasmon resonance (SPR) of gold was observable at 524 nm for oleylamine-AuNPs. In contrast, after the thiolation reaction, SPR of AzBT-AuNPs was observed at 545 nm while the absorption corresponding to the AzBT ligands was clearly observed at 338 nm. Successive irradiations at 365 nm then 470 nm clearly produced variations of the absorbance in the ligand region related to the cis-trans isomerization, demonstrating the switching ability of AzBT-AuNPs in solution. The same experiment was done for NPSAN deposited (see method below) on a quartz substrate. Albeit the signal is weaker and more noisy than in solution, the same behavior is observed in the AzBT ligand region.

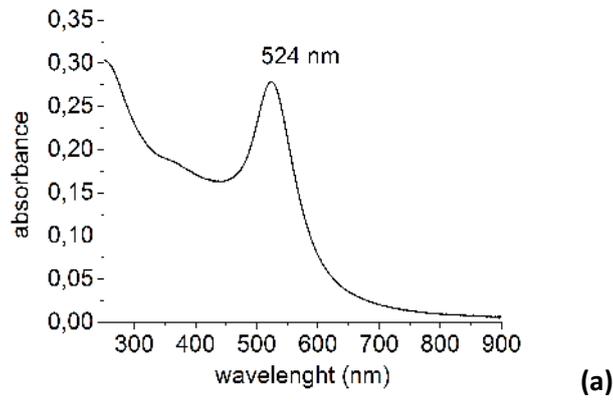

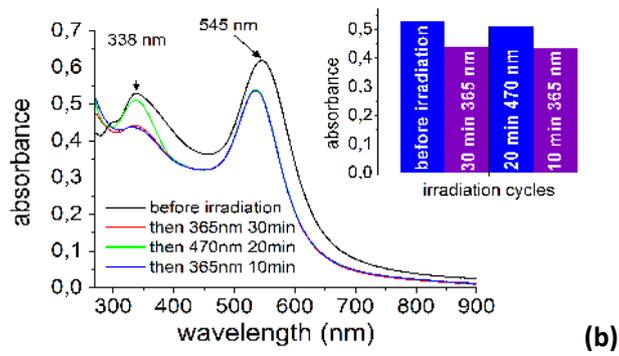

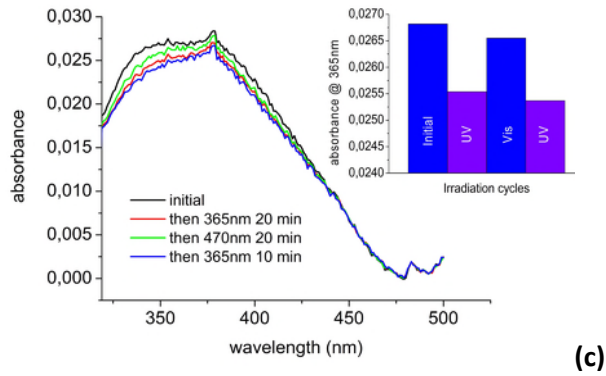

*Figure S2*: *(a): UV-vis spectrum of oleylamine-AuNPs (a) and AzBT-AuNPs (b) in DMF after 2 cycles of UV irradiation at 365 nm then 470 nm. The switching effect is revealed in solution by the absorbance measurement at 340 nm on the spectrum. (c) UV-vis spectra (AzBT region) of NPSAN deposited on quartz substrate.*

## Preparation of NPSAN

The preparation of NPSAN was realized by the Langmuir technique following the method of Santhanam.[6] 100 µL of the previous AzBT-AuNPs solution in TCE were spread on a water meniscus delimited in a pierced teflon petri dish (hole diameter: 2 cm, see Figure S2). Solution in CHCl$_3$ was also possible but as NPs are less soluble, the transfer of NPs in this solvent should be performed immediately before the preparation of the film. Once the half of TCE evaporated, we observed that addition of hexane (~50 µL) favored the formation of the NPSAN. After complete evaporation, the floating film was deposited on a clean substrate. Initially, the Langmuir Schaefer technique (horizontal transfer of the film) was attempted by using an elastomeric PDMS stamp to develop a printing process but results were unsuccessful (only partial transfer was obtained). Therefore, the transfer of the floating film was subsequently realized by dip coating directly on a lithographed substrate.

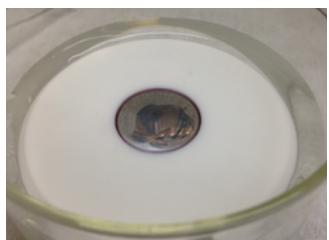

*Figure S3.* AzBT-AuNPs floating film on water surface in a Teflon dish.

## XPS measurements

X-ray photoemission spectroscopy (XPS) experiments were performed to analyze the chemical composition of the NPSAN. AzBT-AuNPs film was deposed on a freshly cleaned silicon wafer (no electrode) and were analyzed by XPS, using a Physical Electronics 5600 spectrometer fitted in an UHV chamber with a base pressure of about $3.10^{-10}$ Torr. The X-ray source was monochromatic Al Kα (hν = 1486.6 eV) and the detection angle was 45° with respect to the sample surface normal. The intensities of XPS core levels were measured as peak areas after standard background subtraction according to the Shirley procedure.[7] The C1s peak was observed at binding energy of 284.3 eV in good agreement with our previous work on AzBT SAM on plane gold surface.[8] Data on the energies, peak intensity and atomic ratios are summarized in Table S1.

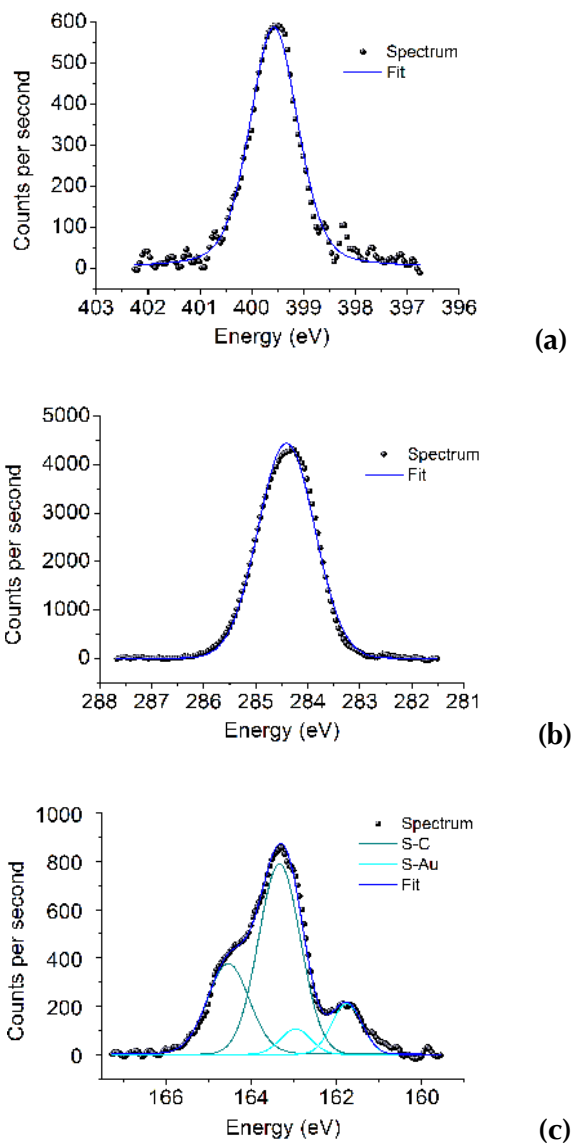

**Figure S4**: XPS spectra of AzBT NPSAN (oleylamine route NPs) in the (a) C1s, (b) S2p, and (c) N1s regions.

Ratios of experimental atomic concentrations S/C (0.15) and N/C (0.07) were in accordance with expected values (5/26=0.19 and 2/26=0.07, respectively). The N1s and S2p peaks appeared at 399.5 eV and 163.3 eV, respectively. These values are also very close to those obtained on AzBT SAMs.[8] The S2p signal was deconvoluted into a pair of doublets with a 1.2 eV splitting energy centered at 163.91 and 162.33 eV ascribed to sulfur bound to carbon (thiophene) and sulfur bound to gold, respectively. The characteristic binding energy of the s2p3/2 at 161.8 eV is in agreement with what is generally found for organothiols chemisorbed on Au.[6, 9-10] This demonstrates that the grafting of AzBT-SH on AuNPs was successful. No sulfonate species was found at higher energy (around 168 eV) proving that no sulfur oxidation occurred. Following the method of Volkert,[11] we evaluated a ligand density of 2560 AzBT molecules per NP (3.2 molecules/nm$^2$) from the experimental S/Au ratio (0.357) of atomic concentrations.

| Attribution | Binding energy (eV) | Area | FWHM | Atomic ratios |
|---|---|---|---|---|
| $C1s_{total}$ | 284.29 | 20554 | 1.28 | |
| $S2p_{total}$ | 163.27 | 3108 | | $S_{total}/C_{total} = 0.15\ (0.19)$ |
| $S2p_{3/2}$-$Au$ | 161.73 | 331 | 0.94 | S-C/S-Au = 5.1 (4) |
| $S2p_{1/2}$-$Au$ | 162.93 | 165 | 0.94 | $S_{total}/Au_{total} = 0.357$ |
| $S2p_{3/2}$-$C$ | 163.31 | 1729 | 1.13 | |
| $S2p_{1/2}$-$C$ | 164.51 | 828 | 1.13 | |
| $N1s_{total}$ | 399.49 | 1411 | 1.11 | $N_{total}/C_{total} = 0.07\ (0.07)$ |
| $Au4f_{total}$ | 84.80, 87.47 | 8689 | | |

**Table S1**: *XPS analysis of AzBT NPSAN deposited on silicon. Areas were corrected by relative sensitivity factors. Experimental values of atomic concentration ratios are compared to theoretical values (in brackets).*

## *Nanogap electrodes*

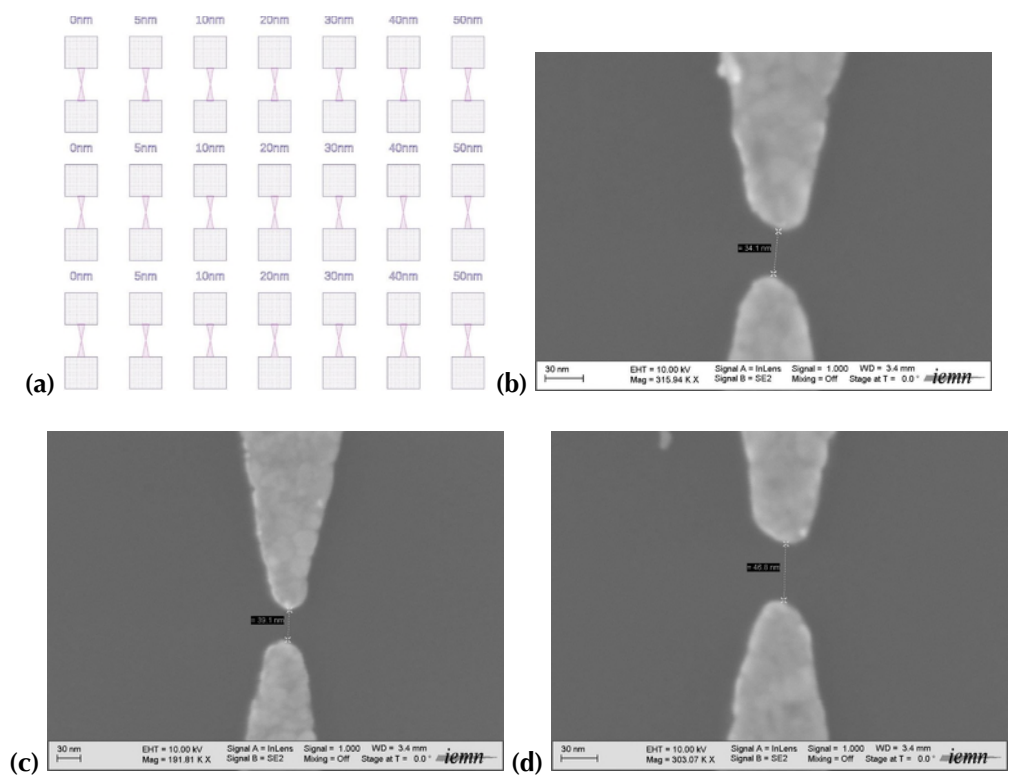

**Figure S5**. (a) Layout of the e-beam mask, SEM image of a some nanogaps before the AzBT-NPSAN deposition: (b) drawn 20 nm, measured 34 nm, (c) drawn 30 nm, measured 39 nm, and (d) drawn 40 nm, measured 47 nm.

## *Calculated optical absorption*

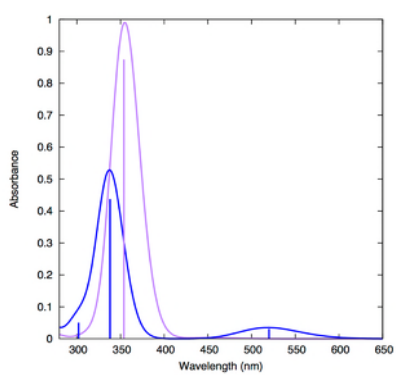

**Figure S6.** Theoretical optical absorption spectra of the AzBT molecule corresponding to the trans-to-cis switching. Calculations performed by TDDFT in Gaussian03 (see main text): purple curve, trans conformation; blue curve, cis conformation.